\documentclass[aps,twocolumn,showpacs,preprintnumbers,amsmath,amssymb,nofootinbib,superscriptaddress,showkeys]{revtex4}

\usepackage{epsfig}
\usepackage{graphicx}
\usepackage{mathptmx}      
\usepackage{latexsym}
\usepackage{epsfig}
\usepackage{graphicx}

\def\1{\'{\i}}
\def\Xint#1{\mathchoice
   {\XXint\displaystyle\textstyle{#1}}%
   {\XXint\textstyle\scriptstyle{#1}}%
   {\XXint\scriptstyle\scriptscriptstyle{#1}}%
   {\XXint\scriptscriptstyle\scriptscriptstyle{#1}}%
   \!\int}
\def\XXint#1#2#3{{\setbox0=\hbox{$#1{#2#3}{\int}$}
     \vcenter{\hbox{$#2#3$}}\kern-.5\wd0}}

\def\dashint{\Xint-}

\def\p{{\bf p}}
\def\k{{\bf k}}
\def\q{{\bf q}}
\def\x{{\bf x}}

\begin{document}

\title{Mass number dependence of BCS pairing gap in the on-shell limit}

\author{E. Ruiz
  Arriola}\email{earriola@ugr.es}
  \affiliation{Departamento de F\'isica At\'omica, Molecular y Nuclear
  and Instituto Carlos I de Fisica Te\'orica y Computacional, Universidad de Granada, E-18071 Granada, Spain}

\author{S. Szpigel}\email{szpigel@mackenzie.br}
\affiliation{Faculdade de Computa\c c\~ao e Inform\'atica,
Universidade Presbiteriana Mackenzie, Brazil}

\author{V. S. Tim\'oteo}\email{varese@ft.unicamp.br}

\affiliation{Grupo de \'Optica e Modelagem Num\'erica (GOMNI), Faculdade de Tecnologia, Universidade Estadual de Campinas - UNICAMP, Brazil}

\date{\today}

\begin{abstract}
The pairing gap plays a fundamental role in the nuclear many-body 
problem and many large scale and accurate mass formula fits suggest
the smooth nuclear mass dependence $\Delta \sim 6(1)~ A^{-1/3}~{\rm MeV}$ in
the liquid drop model which lacks a theoretical motivation. Within the
BCS theory we analyze the impact of phase equivalent interactions on
the pairing gap for a translational invariant many-fermion system such
as nuclear and neutron matter.  To that end we use explicitly the
Similarity Renormalization Group (SRG) transformations.  We show that
in the on-shell and continuum limits the pairing gap vanishes. For
finite size systems the pairing gap can be computed directly from the
scattering phase-shifts by the formula
$$
\Delta_{nn} (p_F)  =  \Delta \epsilon_F ~ \delta^{^1S_0}_{nn}(p_F) /\pi ~ ,
$$ 
where $p_F$ is the Fermi momentum and $\Delta \epsilon_F$ the level spacing 
at the Fermi energy which for the harmonic oscillator shell model becomes 
$\Delta \epsilon_F= \hbar \omega \sim 41 ~ A^{-1/3}~{\rm MeV}$, so that 
$$
\Delta_{nn} (p_F) \sim 4 ~ A^{-1/3}~{\rm MeV} ~ .
$$
The comparison with double differences from binding energies of stable
nuclei is satisfactory and the discrepancy with the large scale
analysis may be attributed to the lack of three-body forces. Nevertheless,
the on-shell two-body interaction provides a basis for the $c~A^{-1/3}$ dependency 
and accounts for 75\% of the coefficient $c$. 
\end{abstract}

\maketitle

\section{Introduction}

The microscopic origin of pairing in nuclei was first driven by the
analogy to the BCS theory of
superconductivity~\cite{Bohr:1958zz}. Since then, the nature of
pairing correlations has provided a lot of insight in Nuclear Physics
providing the rationale for the observed odd-even staggering in
nuclear masses and binding
energies~\cite{bohr1998nuclear,ring2004nuclear} (for a review see
e.g. Ref.~\cite{Dean:2002zx} and references therein). A renaissance of
the subject was experienced by the production of heavy $N \sim Z$
nuclei which are achieved by Radioactive Ion Beams~\cite{Satula:2005fy}
where a pressing need for predicting nuclear masses has also
triggered a re-analysis of nuclear mass formula.

For finite nuclei the pairing gap is defined as a three-point finite
difference
\begin{eqnarray}
\Delta &=& E_{\rm pair} (N,Z) - \frac12 \left[E_{\rm pair} (N+1,Z) + E_{\rm pair} (N-1,Z) \right] \; .\nonumber \\
\label{eq:massdiff}
\end{eqnarray}
The experimental values for nuclei in the range $A=40-208$ with an odd
number of neutrons provide a range of values $\Delta = 0.6-2 ~ {\rm
  MeV}$.  Of course, one cannot directly extrapolate from here the
pairing gaps for nuclear or neutron matter where $N \to \infty$.

Within the liquid drop model framework, the  analysis of
pairing interactions has a long history. The textbook semiempirical
mass formula contains a term already proposed by Bohr and
Mottelson~\cite{bohr1998nuclear} and which after the error analysis of
Ref.~\cite{Toivanen:2008im} reads
\begin{eqnarray}
E_{\rm pair} (N,Z) = \frac{a_P}{2~ A^\frac12} \left[ (-1)^N + (-1)^Z \right] \; ,
\end{eqnarray}
where $a_P= 11.36(2)~ {\rm MeV}$ which for $A=208$ yields $\sim 0.6~
{\rm MeV}$.

\begin{figure}[t]
\begin{center}
\includegraphics[width=8cm]{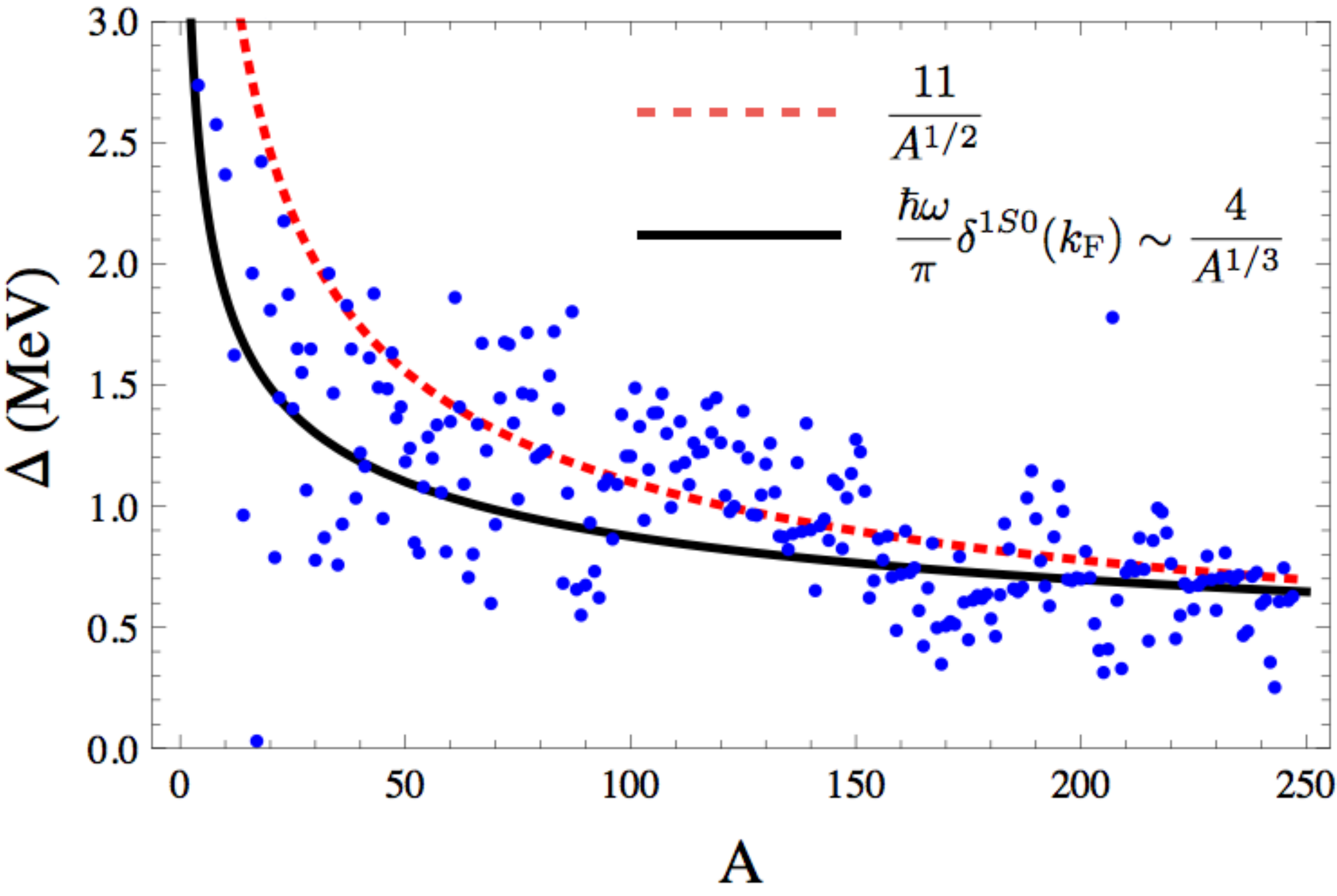}
\end{center}
\caption{BCS pairing gap for the $^1S_0$-state (in ${\rm MeV}$) obtained from
the on-shell phase-shift formula. We compare with double-differences
from stable nuclei for $ 16 < A < 250$. We also show the conventional
 semiempirical mass formula yields $\Delta= 11~A^{-1/2}$.
}
\label{fig:pairing-nuclei}
\end{figure}

Of course, the liquid drop model parametrization represents an
average value, and attempts to obtain the functional mass dependence
of the pairing gap {\it directly} from the mass differences in
Eq.~(\ref{eq:massdiff}) are hindered by the many existing fluctuations
(see e.g. Fig.~\ref{fig:pairing-nuclei}) which need to be disentangled
before a clear mass number dependence may confidently be extracted.

Actually, the original
formula embodying the $A^{-1/2}$ dependence has been deprecated in
favor of a $A^{-1/3}$ dependence when, instead, nuclear mass formulas
including shell and deformation effects are fitted to a large body of
over $2000$ nuclei with a mean standard deviation in the range
$0.2-0.5~{\rm MeV}$. In Section~\ref{sec:mass} we review several
comprehensive analysis in favor of this dependence.

The dynamical origin of this persistent $A^{-1/3}$ dependence of the
nuclear pairing gap remains to date and to our knowledge a mystery. In
the present work we will give theoretical arguments supporting this
behavior and will provide a simple formula where this dependence
arises quite naturally within the conventional BCS theory directly in
terms of the nucleon-nucleon ($NN$) scattering phase-shift, as shown in
Fig.~\ref{fig:pairing-nuclei}.

Our main idea is to show that within the conventional BCS theory and
for a finite but large nucleus the pairing gap can be interpreted as
the energy shift at the Fermi surface due to the $NN$ interaction. This
energy shift can be related for a large system to the $NN$ scattering
phase-shift~\cite{Fukuda:1956zz,DeWitt:1956be}.

The basic ingredient of the BCS equation for the pairing gap is the $NN$
interaction characterized by a potential. Therefore in
Section~\ref{sec:eff-BCS} we review the BCS equation in terms of
effective $NN$ interactions in a translationally invariant system such
as nuclear or neutron matter, restricting our study to the simplest
and most important $^1S_0$ channel. In Section~\ref{sec:grid} we
specify our numerical momentum grid and illustrate a few examples,
including realistic potentials as well as a separable toy model which
will be helpful in order to illustrate some properties of the
solution. While this momentum grid is usually viewed as an auxiliary
tool to solve the BCS equations in practice, the discretization can
also be regarded as a way to impose the natural momentum quantization
in a finite size system.

Traditionally, $NN$ interactions are inferred from the $NN$-scattering
data analysis, and in the particular case of the $^1S_0$ channel from
the corresponding phase-shift. This requires solving the
Lippmann-Schwinger (LS) equation for the potential $V$. However, this does
not determine the $NN$ potential uniquely since one may perform a unitary
transformation $V \to U V U^\dagger$ preserving exactly the same phase
shifts~\cite{Srivastava:1975eg}.

This is analyzed in the interesting momentum grid where the scattering
problem is reformulated equivalently in terms of matrix
diagonalization and energy eigenvalues.  In Section~\ref{sec:on-shell}
we discuss the BCS equation on the grid for unitarily equivalent
potentials, and consider the on-shell limit, i.e. the case where the
potential becomes diagonal in momentum space, for which the BCS
equation can be trivially and analytically solved. We present some
numerical results in Section~\ref{sec:num} illustrating this approach
to the on-shell limit by using the one-parameter similarity
transformation. In Section~\ref{sec:finit} we come to the core of our
construction where we choose an optimal momentum grid by matching the
short distance behavior of the scattering wave function to that of a
harmonic oscillator. This allows to justify our final formula
displaying explicitly the $A^{-1/3}$ dependence of the pairing
gap. Finally, in Section \ref{sec:conl} we come to the conclusions.

\section{Mass number dependence of pairing gap}
\label{sec:mass}

In this section we review some of the nuclear mass fits favoring a
$A^{-1/3}$ dependence of the pairing gap. This is not intended to be
exhaustive but rather to show that a large variety of approaches and
schemes merge into a single and unified pattern.

Vogel {\it et al.}  questioned the Bohr-Mottelson
formula~\cite{Vogel:1984jyh} providing instead $\Delta = 7.1 ~
A^{-1/3}~{\rm MeV}$ for the symmetric case $N=Z$. We will refer to
this case hereafter and quote $\Delta$ in ${\rm MeV}$. Soon thereafter it was
found a significant difference between protons and neutrons with
$\Delta_p = 7.55 ~ A^{-1/3}$ and $\Delta_n = 7.36 ~ A^{-1/3}$
respectively~\cite{Jensen:1984uok}. A better description with a mean
standard deviation of $0.18~{\rm MeV}$ is obtained from fourth order
finite differences when comparing the $A^{-1/2}$ vs the $A^{-1/3}$
obtaining $\Delta = 6.2 ~ A^{-1/3}$~\cite{Madland:1988vix,Moller:1992zz} with satisfactory
statistical properties for the residuals. An analysis of nuclear
masses of $1654$ ground state nuclei with $N,Z\ge 8 $ within a
Thomas-Fermi model including several corrections with RMS deviation in the
fit to masses of $0.655~{\rm MeV}$ needs $\Delta= 6.03 ~ A^{-1/3}$~\cite{Myers:1995wx}. A direct analysis for $N\approx Z$
nuclei has lead to $\Delta = 6.24 ~ A^{-1/3}$~\cite{Vogel:1998km}.  The
disentanglement between pairing and deformation in the odd-even
staggering of nuclear masses has also revealed the inconsistency of
the $A^{-1/2}$ dependence~\cite{Satula:1998ha}. A proton-neutron
pairing study in a microscopic deformed BCS approach for Ge isotopes
$A=64-76$ confirms better agreement with the $A^{-1/3}$ dependence
than with the traditional $A^{-1/2}$ behavior
\cite{Scaronimkovic:2003xv}. The mass number dependence of nuclear
pairing has also been investigated in \cite{Hilaire:2002yv} using
large scale Hartree-Fock-Bogoliubov calculations finding an average
gap of $\Delta_p = 4.52 ~ A^{-1/3}$ and $\Delta_n = 4.9 \,
A^{-1/3}$. Odd-even mass differences from self-consistent mean-field
theory studies yield $\Delta_p = 4.31 ~ A^{-0.31}$ and $\Delta_n =
4.55 ~ A^{-0.31}$~\cite{Bertsch:2008yc}.  The implementation of a new
pairing term in the Duflo-Zucker formula\cite{MendozaTemis:2009ia}
yields $\Delta_n = 5.0943 ~ A^{-1/3}$. Following the pairing form of
this reference, several determinations based on different schemes
essentially confirm this value.  In the improved Janecke mass formula
it is found that $\Delta = 4.9644 ~ A^{-1/3}$~\cite{He:2014aaa},
$\Delta = 5.5108 ~ A^{-1/3}$~\cite{Wang:2010dm} or imposing mirror
nuclei constraint in the mass formula $\Delta = 5.4423 ~
A^{-1/3}$~\cite{Wang:2010uk}.  Uncertainties have been evaluated,
providing $\Delta = 5.396 (137) ~ A^{-1/3}$~\cite{Qi:2014cqa}.  An
improved nuclear mass formula with a unified prescription for the
shell and pairing corrections using explicitly BCS theory obtains
$\Delta = 7.5 ~ A^{-1/3}$ with a remarkable gaussian distribution of
residuals~\cite{Zhang:2014txa}.  Direct fits to four point difference
formulas are compatible with a variety of mass number dependencies, but
when the $A^{-1/3}$ behavior is assumed one obtains $\Delta_p=
6.26(4)~ A^{-1/3}$ and $\Delta_n= 6.22(4) ~
A^{-1/3}$~\cite{Ishkhanov:2014tra}. The approach for the fluctuations
based on a Fourier decomposition produces $\Delta= 5.9 ~
A^{-1/3}$~\cite{Bhagwat:2014jca} and m.s.d.  of $0.266~{\rm MeV}$ for $2353$
nuclei. The neural network approach to the fit including experimental
uncertainties with the previous mass formula in the masses provides
$\Delta = 5.91(25) ~ A^{-1/3}$~\cite{Zhang:2017zvb}. A three-point
fit including a Wigner energy term yields $\Delta= 5.319 (236) ~
A^{-1/3}$ in Ref.~\cite{Cheng:2015sca}.

These studies describe a large body, typically about $2000$, of nuclear
masses with mass formulas containing $10-20$ parameters, including a
pairing term scaling as $A^{-1/3}$ which are determined by least
squares minimization to the experimental nuclear masses database.
While experimental accuracy cannot be obtained, they produce a mean
standard deviations between $0.2-0.5~{\rm MeV}$ for the total
mass. This makes in our view a strong case for this persistent
$A^{-1/3}$ mass number dependence which demands a microscopic
explanation.  For $N=Z$ it can be summarized as in terms of the mean
and standard deviation of all the  studies  cited above
\begin{eqnarray}
\Delta \sim 6(1) ~ A^{-1/3}~ {\rm MeV} \, .
\label{eq:pg-phen}
\end{eqnarray}
The $15 \%$ uncertainty reflects the disparity of the methods and most
likely can be reduced by a judicious weighting, so that it can be
interpreted as a systematic error, but it does show the robustness of
all the approaches as a whole.

In this work we will provide such a microscopic explanation within the
conventional BCS theory taking advantage of the freedom in the
definition of the $NN$ interaction and also the fact that there is a
natural discretization of momenta reflecting the finite size of the
atomic nucleus. Obviously, we expect the continuum limit to correspond
to a translationally invariant system.

\section{Effective $NN$ interactions and BCS equations in Nuclear Matter}
\label{sec:eff-BCS}

In this section we review the basic elements of BCS theory in a
translationally invariant system such as nuclear or neutron matter to
fix our notation (see
e.g. Refs.~\cite{bohr1998nuclear,ring2004nuclear,Dean:2002zx} for more
details). The BCS state provides a paring gap given by
\begin{eqnarray}
\Delta (\k) = -\frac12 \int \frac{d^3 \p}{(2\pi)^3} \frac{V(\k,\p) \Delta(\p)}{E(\p)} \; ,
\end{eqnarray}
where $E(\p)^2 = [(\p^2-p_F^2)/(2M)]^2 + \Delta(\p)^2$ ($M$ is
  the nucleon mass) and the normalization conventions for the three dimensional LS equation are
\begin{eqnarray}
T(\k,\p)=V(\k,\p)+\int \frac{d^3 \q}{(2\pi)^3} \frac{V(\k,\q)
T(\q,\p)}{\p^2/2\mu -q^2/2 \mu} \; ,
\end{eqnarray}
where $2 \mu=M$, i.e. for a local potential
\begin{eqnarray}
V(\k,\p) = \int d^3 \x V(\x) e^{-i (\k-\p) \x} \; .
\end{eqnarray}
After the partial-wave (PW) decomposition,
\begin{eqnarray}
V^S (\p',\p) = \frac{4\pi^2}{M}
\sum_{JMLL'} {\cal Y}_{LS}^{JM} (\hat p') V_{LL'}^{JS} (p',p)
{{\cal Y}_{L'S}^{JM}}^\dagger (\hat p) \; ,
\end{eqnarray}
and
\begin{eqnarray}
\Delta^S (\p) = \sum_{JML} {\cal Y}_{LS}^{JM} (\hat p) \Delta_{L}^{JS} (p) \; ,
\end{eqnarray}
we have~\cite{Dean:2002zx},
\begin{eqnarray}
\Delta_L^J (p) = -\frac1{\pi} \int_0^\infty p^2 dp
\sum_{L'} \frac{V_{L,L'}^J (k,p) \Delta_{L'}^J (p)}{M E(p)} \; ,
\end{eqnarray}
which is the generalized gap equation in all channels.  Here $\Delta$
is in ${\rm fm}^{-1}$ and $V$ is in fm. These equations are solved
iteratively until convergence is achieved. The pairing gap in a given
channel is defined as $\Delta_F=\Delta(p)|_{p=p_F}$.

In this paper we will restrict our analysis to the simplest $^1S_0$ channel
where the LS equation reads (we omit channel indices)
\begin{eqnarray}
T(k',k) = V(k',k) + \frac{2}{\pi} \int_0^\infty dq
V(k',q) \frac{ q^2 }{k^2-q^2} T(q,k) \; ,
\label{eq:LS}
\end{eqnarray}
The imaginary part of the T-matrix, ${\rm Im}
T(k',k)$, is fixed by unitarity. Passing to the reaction matrix defined by the real part $K(k',k)
= {\rm Re}\, T(k',k)$ we get
\begin{eqnarray}
K(k',k) = V(k',k) + \frac{2}{\pi} \dashint_0^\infty dq
V(k',q) \frac{ q^2 }{k^2-q^2} K(q,k) \; ,
\label{eq:LS-K}
\end{eqnarray}
whence the phase shift reads
\begin{eqnarray}
\tan \delta(k) = -\frac{K(k,k)}{k} \; .
\label{eq:phase}
\end{eqnarray}
On the other hand, for the $^1S_0$ channel the BCS equation becomes
\begin{eqnarray}
\Delta (k) = -\frac1{\pi} \int_0^\infty p^2 dp
\frac{V (k,p) \Delta(p)}{M E(p)} \; ,
\label{eq:BCS}
\end{eqnarray}
Realistic and effective $NN$ interactions, $V(k'k)$, have been used to
compute the pairing gap for nuclear and neutron matter in several
schemes~\cite{Dean:2002zx}. Most often the BCS approach is based on
having the scattering phase-shift as the basic input of the
calculation. On the other hand, there is an arbitrariness in this
procedure, as there are infinitely many interactions leading to the
identical phase-shift. In this paper we analyze these ambiguities.

For instance, in the case of the pairing gap in the $^1S_0$ channel,
most $V_{\rm low-k}$ calculations (not surprisingly) provide a BCS gap
which has maximum at Fermi momentum $p_F \sim 0.8 ~ {\rm fm}^ {-1}$
and strength about $3 ~ {\rm
  MeV}$~\cite{Sedrakian:2003cc,Hebeler:2006kz}. Chiral pion dynamics
has been advocated in Ref.~\cite{Kaiser:2004uj} with similar
results. We consider this as an educated guess for the total gap
including polarization and short-distance correlations since {\it ab
  initio} calculations may provide completely different
results~\cite{Gezerlis:2009iw,Gandolfi:2009tv,Baldo:2010du}. It is,
however, disconcerting that the BCS gap is so different and so much
scheme dependent.

There are claims in the literature that what determines the pairing
gap are the phase-shifts~\cite{Elgaroy:1997ti} and finite nuclei
calculations are carried out in
Ref.~\cite{Baroni:2009eh,Idini:2011gm}. We will see that this is not
strictly true. In medium $T$-matrix has been used to provide an
improvement on the standard BCS theory~\cite{Bozek:2002ry} yielding a
30\% reduction in the $^1S_0$ pairing gap.

\begin{figure}[t]
\begin{center}
\includegraphics[width=8cm]{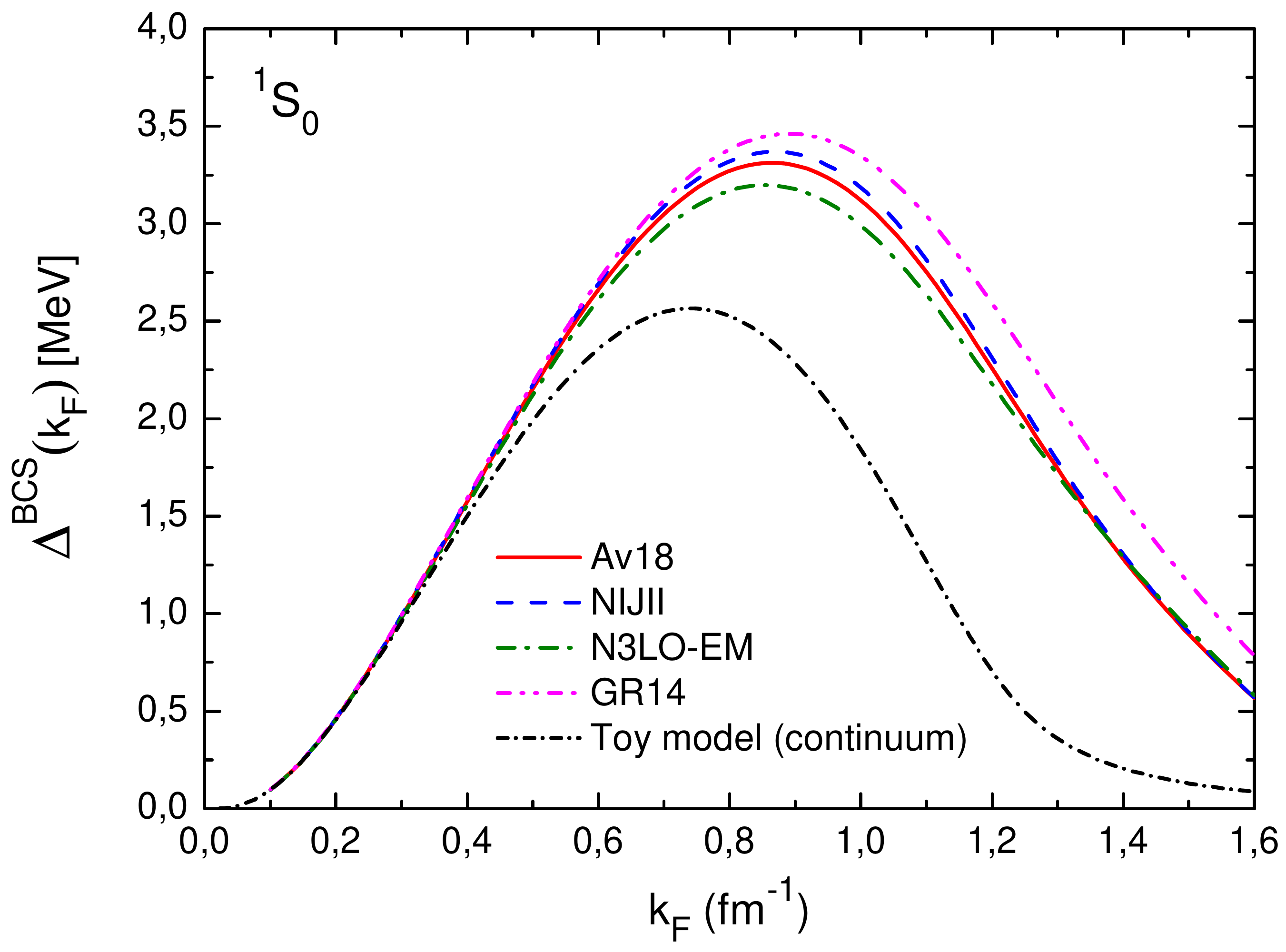}
\end{center}
\caption{BCS pairing gap for the $^1S_0$-state in ${\rm MeV}$ of the toy model
  compared to realistic potentials AV18~\cite{Wiringa:1994wb},
  NijII~\cite{Stoks:1994wp}, N3LO-EM~\cite{Entem:2003ft} and
  GR14~\cite{Perez:2014yla}, as a function of the Fermi momentum $k_F$
  (in ${\rm fm}^{-1}$).}
\label{fig:BCS-toy-vs-pots}
\end{figure}

The previous equations neglect three- and many-body forces (see
however Ref.~\cite{Holt:2013vqa,Furnstahl:2013oba}). While this may be
a crude assumption, let us remind that disentanglement of three-body
forces relies on the off-shellness of the interaction which can
actually be traded for three-body forces. Thus, we will restrict
ourselves to BCS many-body wave functions in a first approximation and
will judge on the need of three-body forces at the end.

\section{BCS pairing equations on a momentum grid}
\label{sec:grid}

The BCS equations can be solved numerically on an $N$-dimensional
momentum grid, $p_1 < \dots p_N $ ~\cite{Szpigel:2010bj} by
implementing a high-momentum ultraviolet (UV) cutoff, $p_{\rm
  max}=\Lambda$, and an infrared (IR) momentum cutoff $p_{\rm min} =
\Delta p$. The integration rule becomes
\begin{eqnarray}
\int_{\Delta p}^\Lambda dp f(p) \to \sum_{n=1}^N w_n f(p_n) \, .
\end{eqnarray}
The BCS equations on the grid follow from inserting the completeness
relation in discretized momentum-space
\begin{eqnarray}
1=\frac{2}{\pi}\sum_{n=1}^N w_n p_n^2 | p_n \rangle \langle p_n | \, ,
\end{eqnarray}
\noindent
and defining the matrix-element as $V(p_n,p_m) \equiv \langle p_n |
V | p_m \rangle $.  For instance, the eigenvalue problem on the grid
may be formulated as (bound states correspond to $P_\alpha= i \gamma$)
\begin{eqnarray}
H \varphi_\alpha (p) = P_\alpha^2 \varphi_\alpha (p)   \, ,
\label{eq:eigenv}
\end{eqnarray}
where the matrix representation of the hamiltonian reads
\begin{eqnarray}
H(p_n,p_m) = p_n^2 \delta_{n,m} + \frac{2}{\pi} w_n p_n^2 V(p_n,p_m)  \, .
\label{eq:hamil}
\end{eqnarray}
Let us consider the BCS pairing gap equation on the grid
\begin{eqnarray}
\Delta (p_n) = -\frac{2}{\pi} \sum_{k=1}^N w_k p_k^2 \frac{V(p_n,p_k)\Delta (p_k)}{2 ME (p_n)} \;,
\end{eqnarray}
where $ 2 M E (p_n) = \sqrt{(p_n^2-p_F^2)^2+ 4 M^2 \Delta
  (p_n)^2}$. Of course, for consistency the Fermi momentum must also
belong to the grid, $p_F= p_m$, so $ 2 M E (p_n) =
\sqrt{(p_n^2-p_m^2)^2+ 4 M^2 \Delta (p_n)^2}$. We use an iterative
method of solution (for several strategies see
e.g.~\cite{khodel1996solution}).

Solutions for the realistic potentials AV18~\cite{Wiringa:1994wb},
NijII~\cite{Stoks:1994wp}, N3LO-EM~\cite{Entem:2003ft} and the more
recent Granada potential (GR14)~\cite{Perez:2014yla} are displayed in
Fig.~\ref{fig:BCS-toy-vs-pots} for a large number of grid points
$N=200$ and a maximum momentum $p_{\rm max}= 30 ~{\rm fm}^{-1}$.

Most analysis based on BCS usually end here. However, we want to
stress that the only physical input information in these calculations
is contained in the phase-shift but {\it not} in the potentials. As it
is well known, this generates some off-shell ambiguity which cannot be
directly related to measurable physical
information~\cite{Srivastava:1975eg}. This off-shellness corresponds
to the non-diagonal matrix elements of the potential $V(p',p)$.  In
fact, one can make a unitary transformation $V \to U V U^\dagger$ such
that phase-shifts remain invariant. In the next sections we will show
that the BCS pairing gap can in principle depend strongly on this
unitary transformation and hence on the off-shellness.

Besides, while the momentum grid is usually regarded as an auxiliary
element for solving the BCS gap equation, we will show that it
actually may be tailored to encode some relevant physical information,
suggesting that in fact finite momentum grids may represent the finite
size of the system. Moreover, we will show that using the inherent
arbitrariness of the off-shellness in the potential one may get a
large variety of results for the pairing gap. As a matter of fact, we
will present a scheme which is free of any off-shell ambiguities, and
for this scheme the continuum limit is shown to produce a vanishing
BCS gap for an infinitely large system, as suggested by the abundant
phenomenological large scale analysis quoted in Section~\ref{sec:mass}
and suggesting $\Delta \sim a_P ~A^{-1/3}$.

\section{Phase equivalent interactions and the on-shell limit}
\label{sec:on-shell}

\def\tr{{\rm Tr}}

Quite generally, for a given hamiltonian we can always perform a
unitary transformation $H \to U H U^ \dagger $ keeping the phase-shift
invariant. On the momentum grid $p_n$ with finite $N$ the definition
of the phase-shift must be specified, since on the one hand one
replaces the scattering boundary conditions with standing waves
boundary conditions and on the other hand one wants to preserve the
invariance under unitary transformations on the grid. As we have
discussed in our previous work~\cite{Arriola:2014fqa} the energy-shift
formula,
\begin{eqnarray}
\delta^{\rm ES} (p_n) = - \pi  \frac{P_n^2-p_n^2}{2 w_n p_n} \; ,
\label{eq:es-ps}
\end{eqnarray}
with $P_n^2$ the $n-$th ordered eigenvalue, see  Eq.~(\ref{eq:eigenv}), of
the grid hamiltonian in Eq.~(\ref{eq:hamil}) provides a suitable
definition for the case where no bound states are
present~\footnote{The case with one bound-state implementing
  Levinson's theorem on the finite momentum grid has been discussed in detail
  in our previous work~\cite{Arriola:2014fqa} in connection to the $^
  3S_1$ channel. This implies a suitable modification of the energy-shift formula. We have also shown there that the LS
  phase-shifts on a finite grid {\it are not} preserved under unitary transformations.}.

The unitary transformation $U$ can be quite general, and for our study
we will generate them by means of the so-called Similarity Renormalization Group (SRG), proposed by Glazek and
Wilson~\cite{Glazek:1993rc,Glazek:1994qc} and independently by
Wegner~\cite{wegner1994flow} who showed how high- and low-momentum degrees of
freedom can decouple while keeping scattering equivalence.

The general SRG equation is given by~\cite{Kehrein:2006ti},
\begin{eqnarray}
\frac{d H_s}{ds} = [[ G_s, H_s],H_s] \; ,
\label{eq:SRG}
\end{eqnarray}
and supplemented with a generator $G_s$ and an initial condition at
$s=0$, $H_0$. This correspond to a one-parameter operator evolution
dynamics and, as it is customary, we will often switch to the SRG
cutoff variable $\lambda=s^{-1/4}$ which has momentum dimensions. The
unitary character of the transformation follows from the trace
invariance property $\tr (H_s)^n= \tr (H_0)^n$ which holds due to the
commutator, and hence $H_s = U_s H_0 U_s^\dagger $. The generator
$G_s$ can still be chosen according to certain requirements, and
three popular choices are the kinetic energy
$T$~\cite{Glazek:1994qc} (Wilson-Glazek generator), the diagonal part
of the Hamiltonian ${\rm Diag}(H)$~\cite{wegner1994flow}(Wegner
generator) or a block-diagonal (BD) generator $P H_s P + Q H_s Q $
where $P+Q=1 $ are orthogonal projectors $P^2 = P$, $Q^2 = Q$ ,
$QP=PQ=0$, for states below and above a given
momentum scale~\cite{Anderson:2008mu}.

On the finite momentum grid the SRG equations become a set of
non-linear coupled equations. For the Wegner generator, which will be
taken here for definiteness, the equations take a quite simple form
\begin{eqnarray}
\frac{d H_s (p_n,p_m)}{ds} &=& \frac2{\pi} \sum_{k}  H_s (p_n,p_k) w_k p_k^2 H_s (p_k,p_m) \nonumber \\ &\times& \left[H_s (p_n, p_n)+ H_s (p_m,p_m)
- 2 H_s (p_k, p_k)  \right] \; .  \nonumber \\
\end{eqnarray}
%


The fixed points of the SRG evolution with a given generator $G_s$ correspond to the stationary solutions of the SRG flow equations for the matrix-elements of the hamiltonian,
\begin{equation}
\frac{d H_s (p_n,p_m)}{ds} = \langle n| [[G_s,H_s],H_s]|m \rangle=0 \; ,
\end{equation}
\noindent
which implies that in the infrared limit $s \to \infty$ ($\lambda \to 0$) the hamiltonian $H_s$ becomes diagonal, i.e.
\begin{equation}
\lim_{\lambda \to 0} H_\lambda (p_n,p_m)= P_n^2 \delta_{nm} \; ,
\end{equation}
\noindent
In this limit the potential $V_\lambda (p_n,p_m)$ also becomes diagonal, and the LS equation reaction matrix coincides with the
potential. Thus, all off-shellness is eliminated in the SRG infrared
limit $\lambda \to 0$. An important result derived in
Ref.~\cite{Arriola:2014fqa} is
\begin{eqnarray}
\delta_n^{G} = - \pi \lim_{\lambda \to 0} \frac{H_{n,n}^{G,\lambda}-p_n^2}{2 w_n p_n} \; .
\end{eqnarray}

For this phase equivalent hamiltonian family, the BCS equation can be written as
\begin{eqnarray}
\Delta_\lambda (p_n) = -\sum_{k=1}^N \frac{[H_\lambda (p_n,p_k) - p_n^2 \delta_{nk}] \Delta_\lambda (p_k)}{2 ME_\lambda (p_n)} \; ,
\end{eqnarray}
where $ 2 M E_\lambda (p_n) = \sqrt{(p_n^2-p_F^2)^2+ 4 M^2 \Delta_\lambda (p_n)^2}$.
Clearly, the BCS pairing gap becomes a
function of the SRG parameter $\lambda$, {\it without} ever changing
the phase-shifts. Actually, in the infrared limit $\lambda \to 0$ the Hamiltonian becomes diagonal and hence we get
\begin{eqnarray}
\Delta_0 (p_n) = -\frac{P_n^2-p_n^2}{2M} \frac{\Delta_0 (p_n)}{E_0 (p_n)} \; ,
\end{eqnarray}
where the notation $\Delta_0 (p_n) \equiv \lim_{\lambda \to 0} \Delta_\lambda (p_n)$ has been introduced.
The solution is non trivial $\Delta_0 (p_n) \neq 0 $ provided
\begin{eqnarray}
1 = -\frac{P_n^2-p_n^2}{2M E_0 (p_n)} \; ,
\end{eqnarray}
and taking the grid point to be the Fermi momentum $p_m = p_F$ we get $E_0(p_m) = \Delta_0(p_m)$ and hence
\begin{eqnarray}
\lim_{\lambda \to  0} \Delta_\lambda (p_m) = \lim_{\lambda \to 0}
-\frac{H_\lambda (p_m,p_m)-p_m^2}{2M} = w_m \frac{p_m \delta^{\rm ES} (p_m)}{\pi M} \; .\nonumber\\
\end{eqnarray}
Thus, the pairing gap is determined by the energy-shift at the
Fermi surface in the $^1S_0$ channel for on-shell interactions. Not
surprisingly, only the phase-shift appears in the final result. Note that since $\Delta_0 (p_m) \ge 0$ the equation makes
sense only for $\delta (p_m) \ge 0$. Note also, that the integration
weights $w_n$ appear explicitly in the formula, and in the continuum
limit $N \to \infty$ they vanish as $w_n= {\cal O}(1/N)$ as
expected~\footnote{For a large Chebychev grid for instance if the
  location $m$ corresponds to $p_F$ we have $w_F = 4 M p_F $.}. Therefore, if we denote by $\Delta p_F \equiv
w_F $ the integration weight corresponding to the Fermi momentum, in
the continuum limit the BCS pairing gap becomes
\begin{eqnarray}
\Delta_F \sim \Delta p_F \frac{p_F \delta (p_F)}{\pi M}
\label{eq:DeltaPF}
\end{eqnarray}
whenever $\Delta_F > 0$ and zero otherwise. This is our main result.
Note that while the shape is rather universal, the strength is related
to $\Delta p_F$ which ultimately depends on the system size $R$ and
geometry, and for large systems $\Delta p_F= {\cal O}(1/R)$.  While
the simplicity of the result may look as being naive, in the next Section we
will check by explicit numerical calculations, that
Eq.~(\ref{eq:DeltaPF}) is indeed correct.

\begin{figure}[t]
\begin{center}
\epsfig{figure=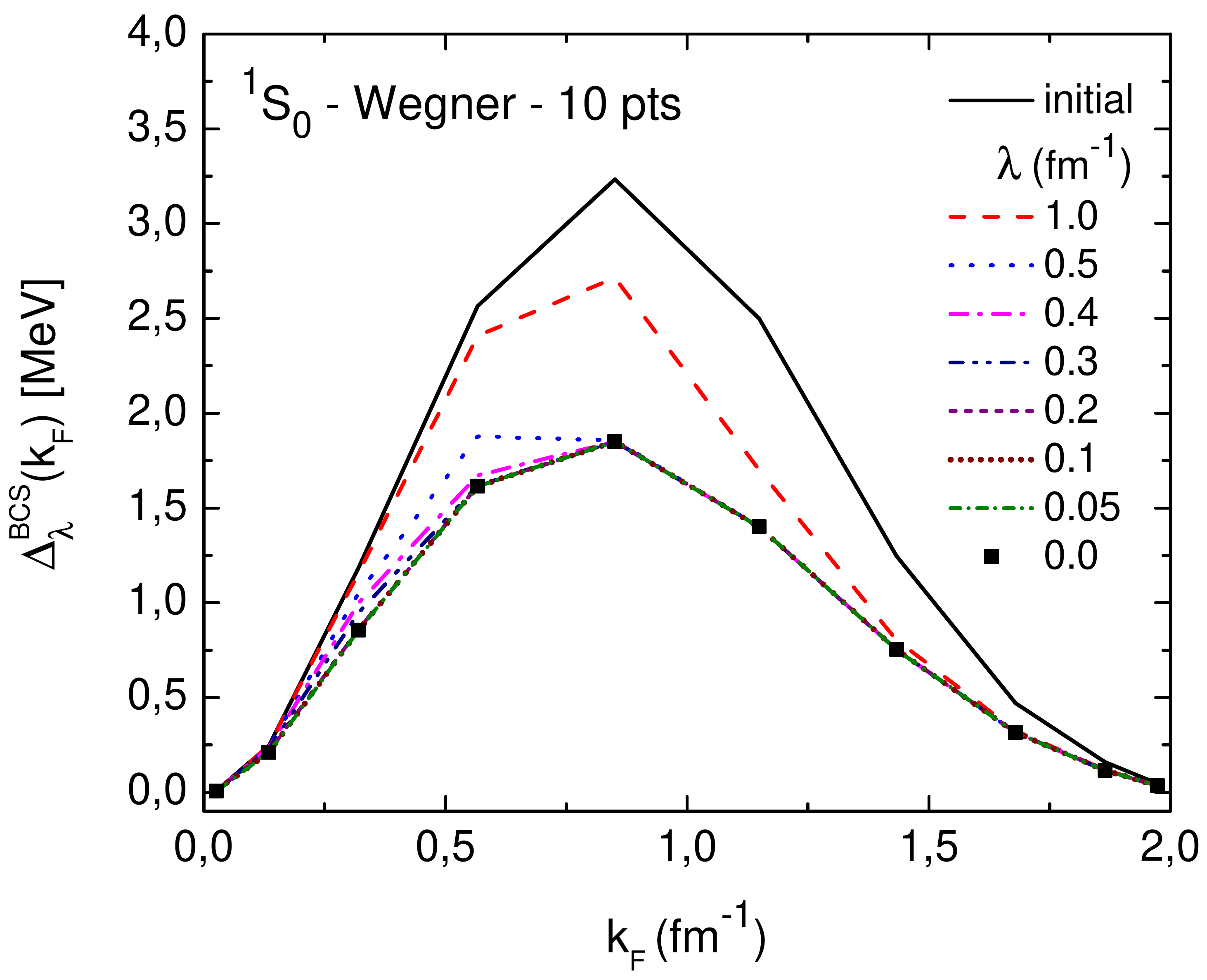,width=0.47\linewidth}
\epsfig{figure=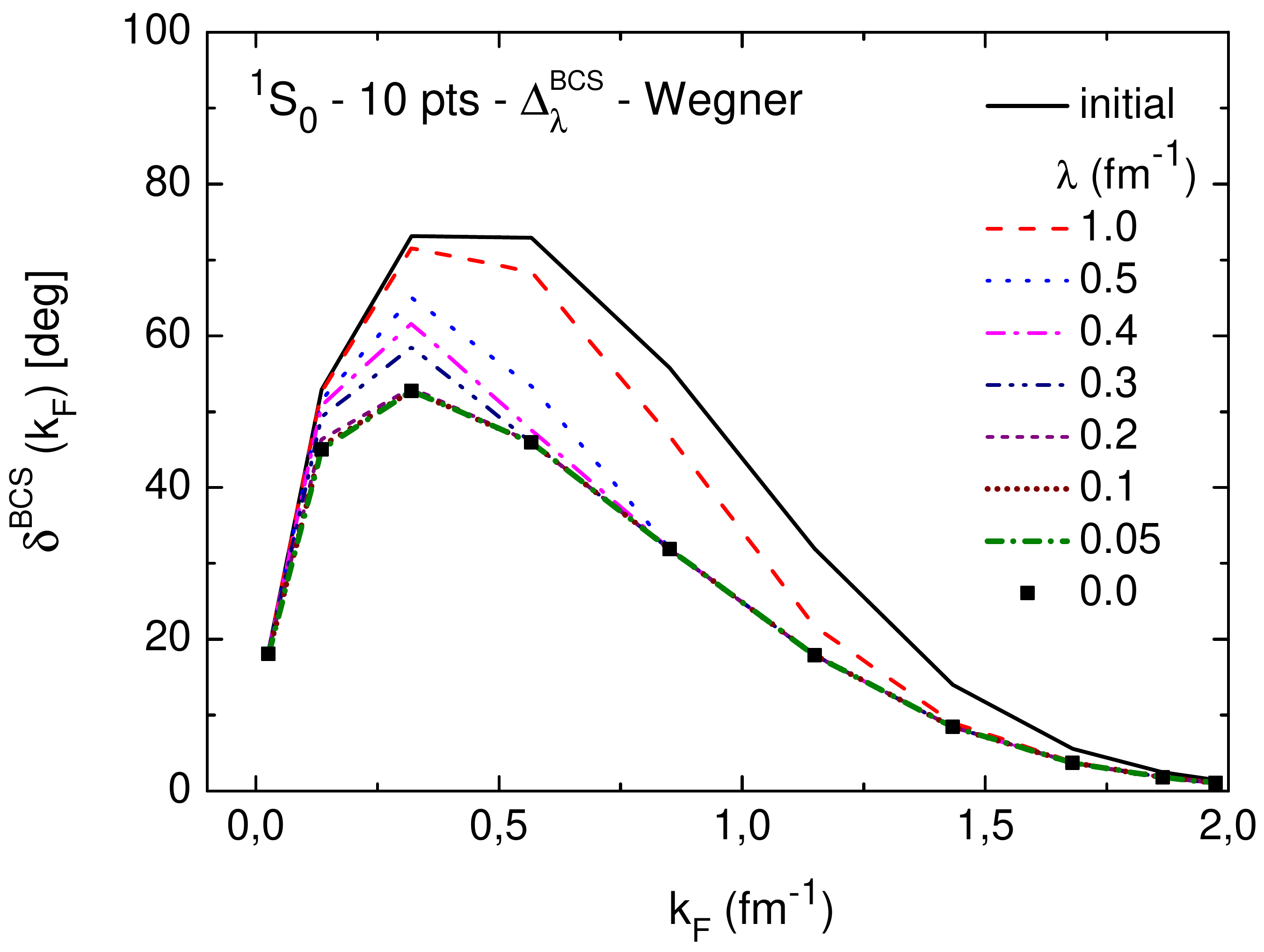,width=0.5\linewidth} \\
\epsfig{figure=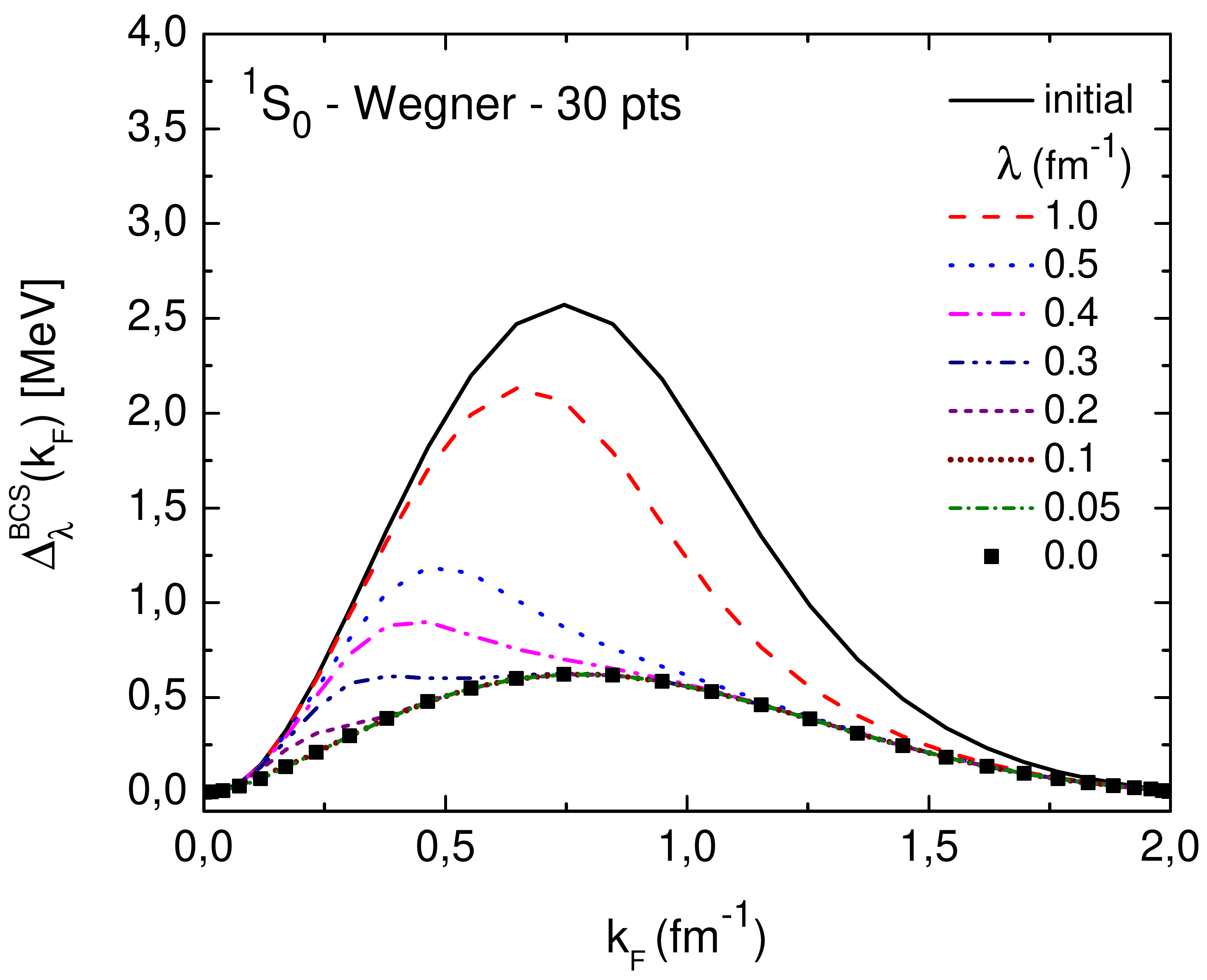,width=0.47\linewidth}
\epsfig{figure=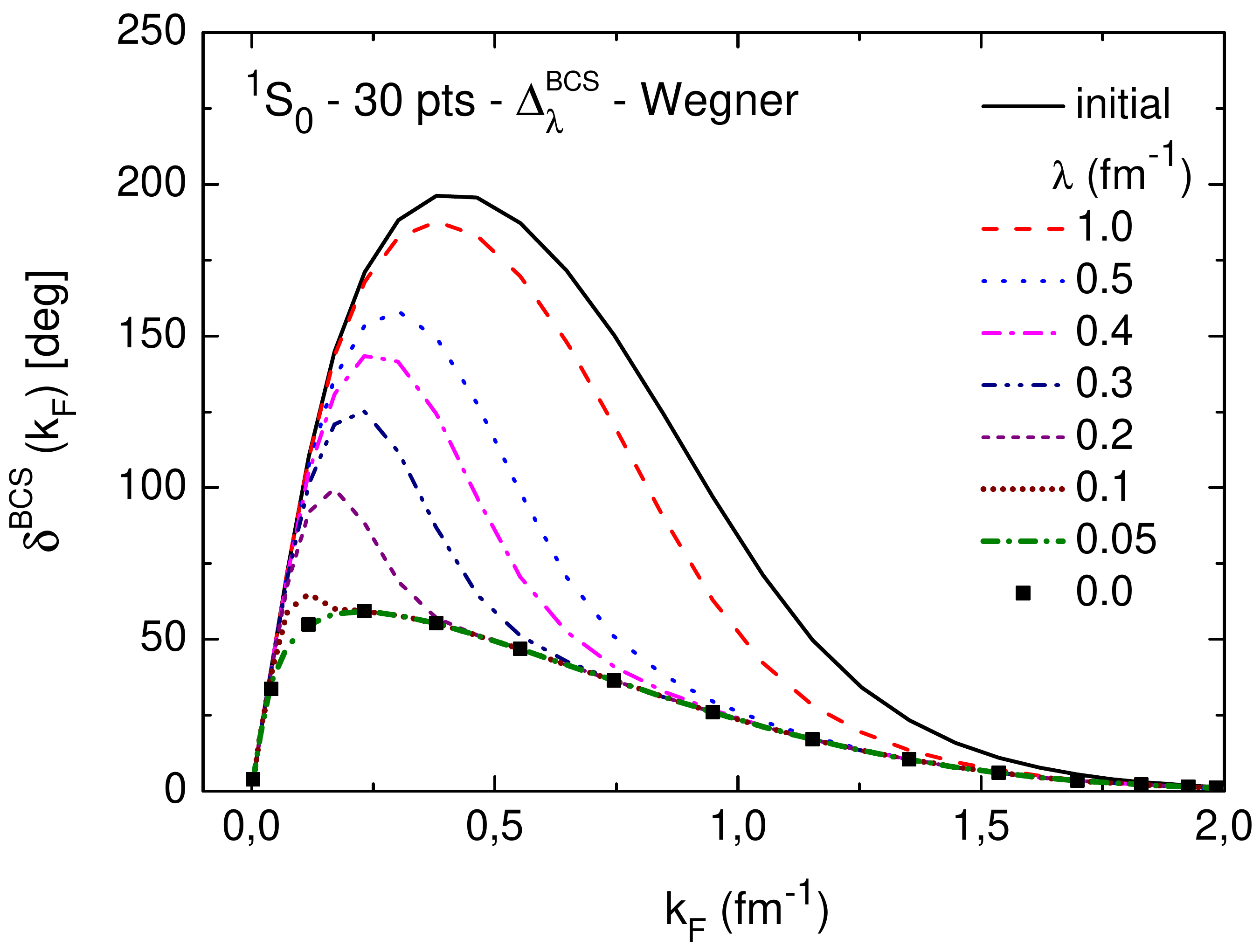,width=0.5\linewidth} \\
\epsfig{figure=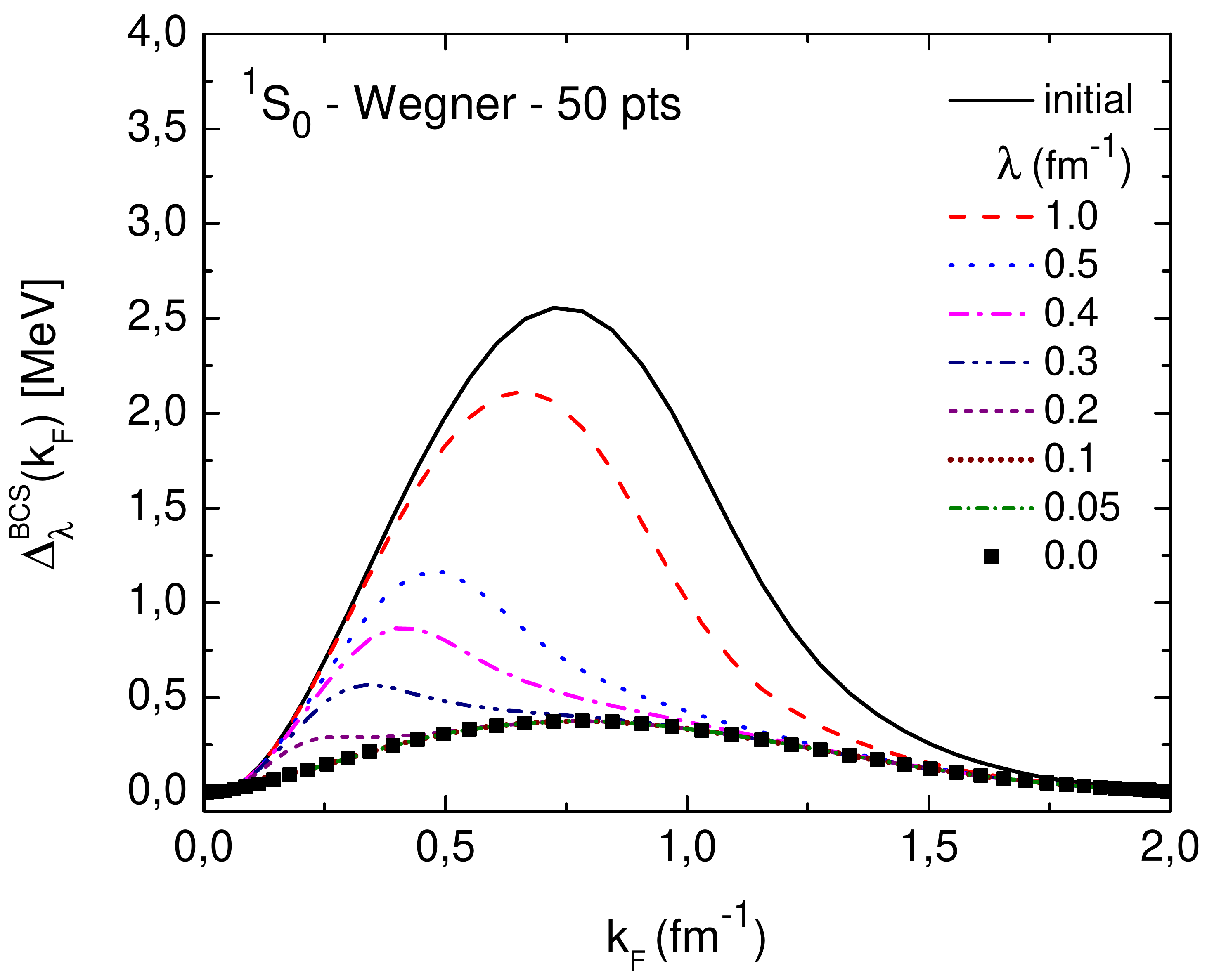,width=0.47\linewidth}
\epsfig{figure=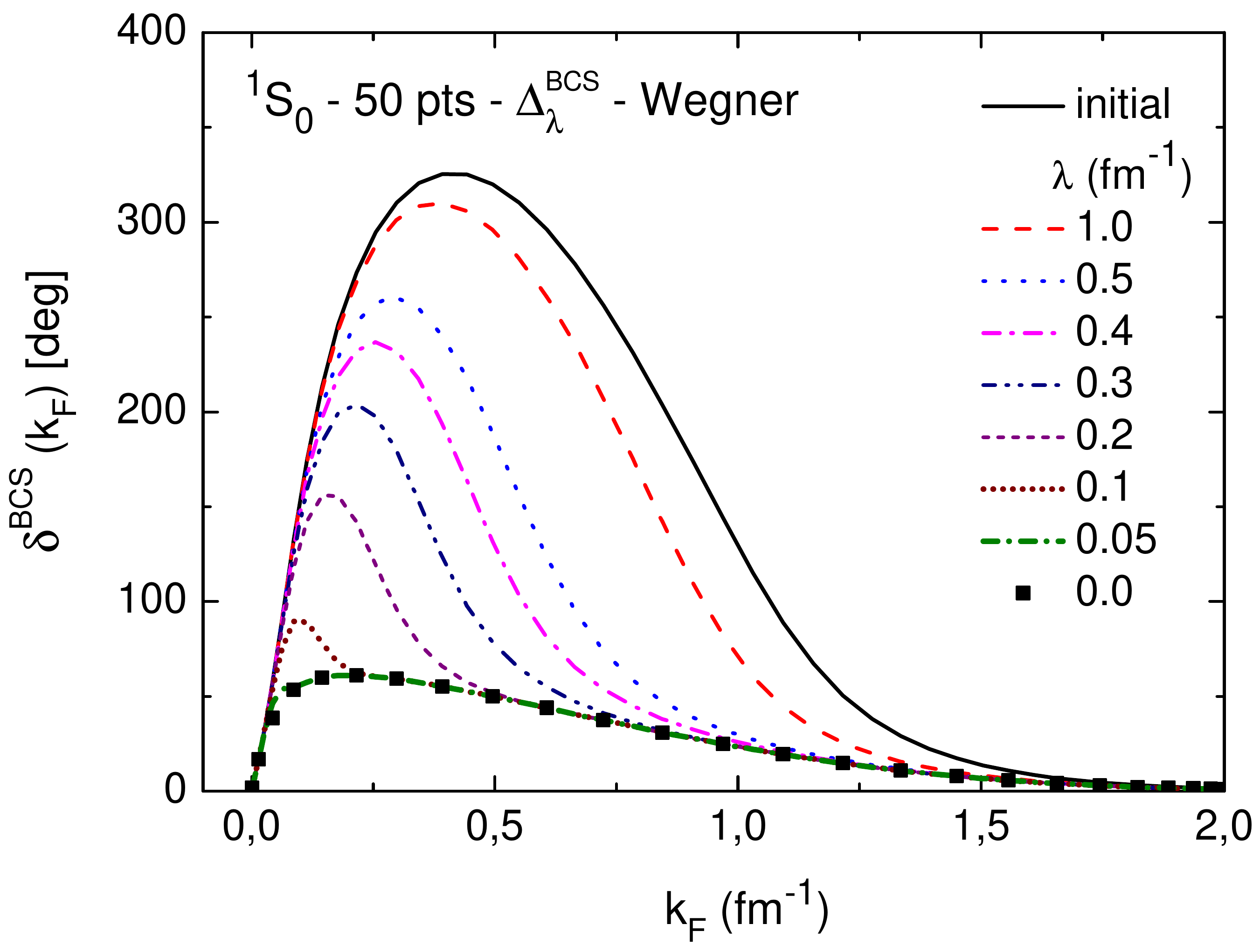,width=0.5\linewidth} \\
\epsfig{figure=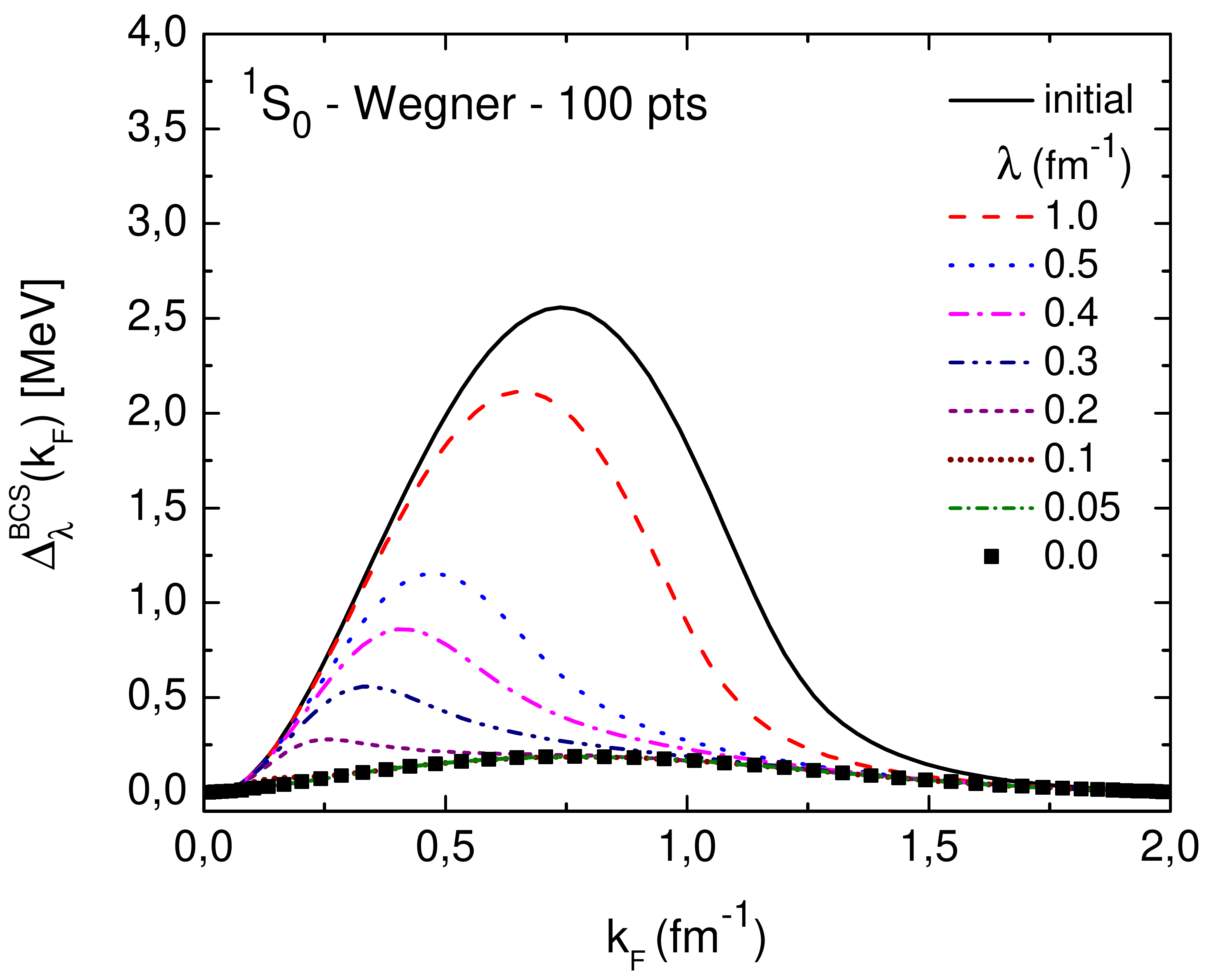,width=0.47\linewidth}
\epsfig{figure=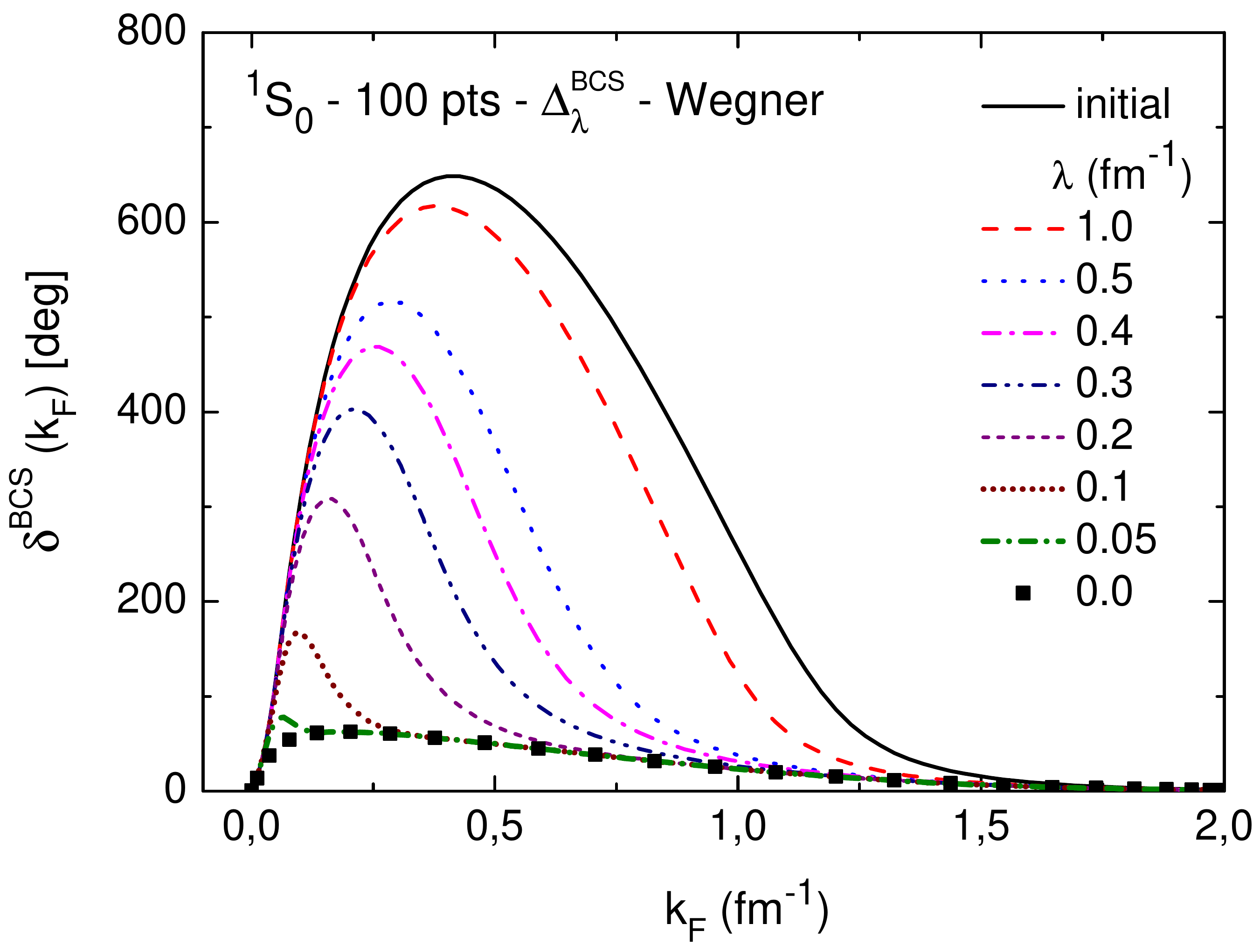,width=0.5\linewidth}
\end{center}
\caption{BCS pairing gap for the $^1S_0$-state in MeV (left panel)
  and the corresponding BCS phase-shift in degrees (right panel) as a
  function of the Fermi momentum $k_F$ (in ${\rm fm}^{-1}$). We show
  the SRG evolution for different values of the SRG cutoff $\lambda$ and for
  different number of grid points. In the left panel we also compare
  with the SRG invariant phase shift deduced as an energy shift.
}
\label{fig:BCS-SRG}
\end{figure}

\section{Numerical Checks}
\label{sec:num}

It is interesting to analyze numerically the behavior of the pairing
gap as a function of the SRG cut-off towards the on-shell limit. These
are demanding calculations particularly with interactions having a
strong short distance repulsive core which provide long momentum
tails. Therefore and for computational reasons, in the present study
of the $^1S_0$ neutron-neutron channel we will use the toy model
gaussian separable potential discussed in our previous
works~\cite{Arriola:2013era,Arriola:2014aia,Arriola:2014fqa,Arriola:2016fkr}
because the long momentum tails are suppressed from the
start~\footnote{The separable potential proposed in
  Ref.~\cite{Elgaroy:1997ti} uses a solution of the inverse scattering
  problem~\cite{Tabakin:1969mr} with the physical phase-shift. We keep
  the gaussian form since the tails are short and as we will see the
  computational effort gets considerably reduced for SRG
  purposes.}. For completeness we review the model in
Appendix~\ref{sec:toy}. The performance of the toy model is rather
good below for the phase-shifts below $p=1 ~{\rm
  fm}^{-1}$~\cite{Arriola:2013era}. The resulting pairing gap can be
obtained quite accurately as depicted in
Fig.~\ref{fig:BCS-toy-vs-pots} and compared to some realistic
potential calculations.  As we see, and for our illustration purposes,
it yields a sufficiently reasonable behavior compared to the realistic
potentials AV18~\cite{Wiringa:1994wb}, NijII~\cite{Stoks:1994wp},
N3LO-EM~\cite{Entem:2003ft} and GR14~\cite{Perez:2014yla}. The good
feature of the toy model is that when using the iterative method of
solution~\cite{khodel1996solution} instead of $N=200$ and $p_{\rm
  max}= 30 ~{\rm fm}^{-1}$ good convergence is achieved for $N=50$ and
$p_{\rm max}= 2 ~{\rm fm}^{-1}$.

In Fig.~\ref{fig:BCS-SRG} we show the evolution of the pairing
gaps $\Delta_\lambda (p_n)$ obtained by solving the BCS equation on
the grid as a function of the SRG cutoff $\lambda$ for the range
$\lambda=1, \dots, 0.05 ~ {\rm fm}^{-1}$ and for several choices on
the number of points $N=10, \dots , 100$.  Alternatively, we
illustrate the scaling behavior of the BCS pairing gap by defining the
``BCS phase-shift'' as
\begin{eqnarray}
\delta_\lambda^{\rm BCS} (p_n) = \frac{ \Delta_\lambda^{\rm BCS} (p_n) \pi M}{w_n p_n} \; ,
\label{eq:BCS-ps}
\end{eqnarray}
which, as expected, converges to the phase-shift obtained from the
energy-shift formula, in the limit $\lambda \to 0$.

We note that along the SRG-trajectory
the phase-shift remains constant if we take the energy-shift
definition~\cite{Arriola:2014aia} as in Eq.~(\ref{eq:es-ps}), i.e.
\begin{eqnarray}
\delta_\lambda^{\rm ES} (p_n) = \delta_\infty^{\rm ES} (p_n) = \delta_0^{\rm ES} (p_n) \; .
\end{eqnarray} Furthermore, we remind that as pointed out and illustrated in
Ref.~\cite{Arriola:2014aia} the LS phase-shift, see
Eq.~(\ref{eq:phase}), {\it does not} fulfill the phase-invariance on
the finite grid, but only in the continuum limit, i.e. for $N \to
\infty$. For reasons that will become clear below we are interested in
having a definition, such as Eq.~(\ref{eq:BCS-ps}), of the phase-shift
which is invariant along the SRG trajectory for {\it any} number of
grid points N.  It is important to emphasize that taking the on-shell
situation corresponds to the infrared limit $\lambda \to
0$~\cite{Arriola:2013gya} as the interaction becomes diagonal.

\begin{figure}[h]
\begin{center}
\epsfig{figure=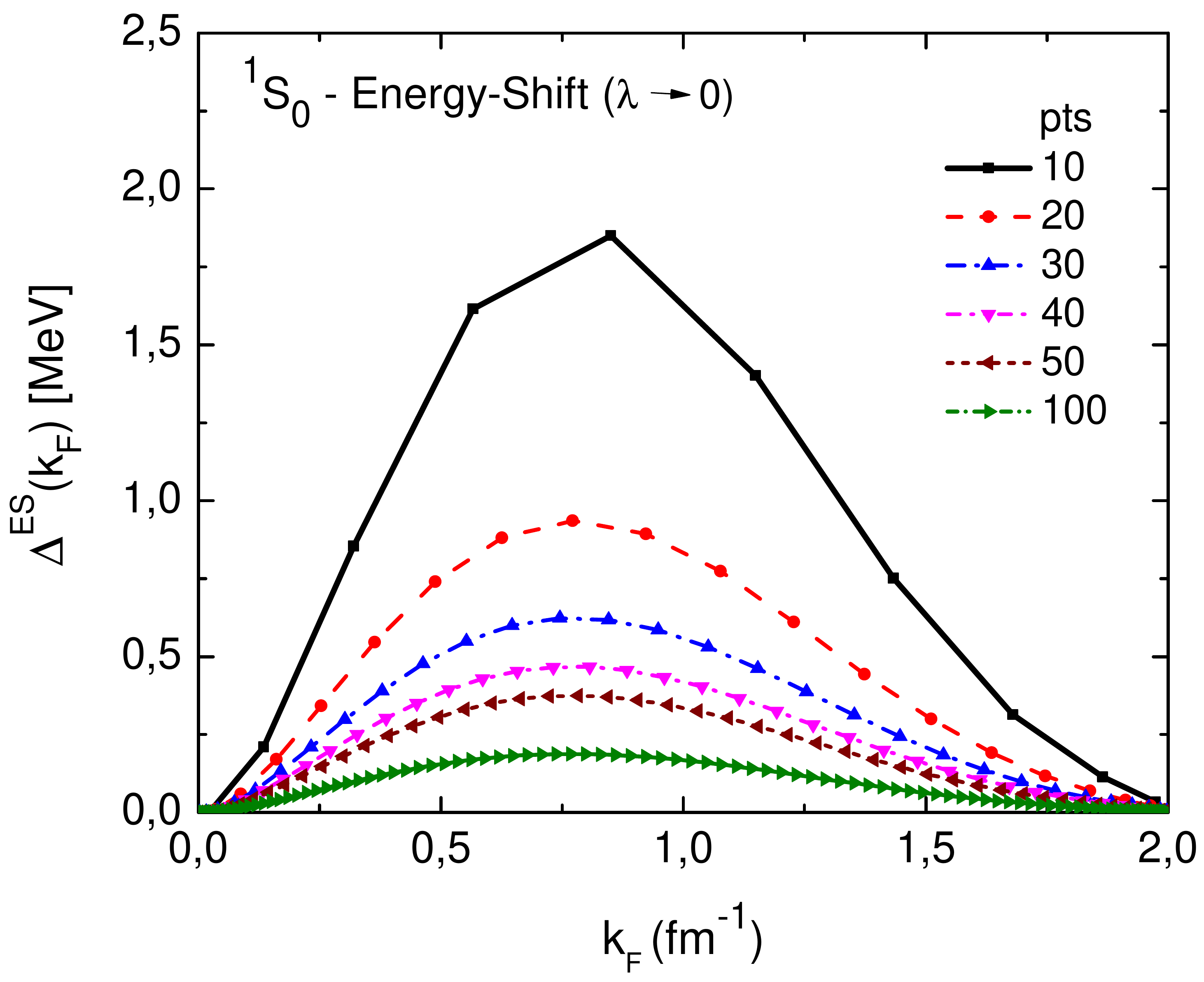,width=0.7\linewidth}\\
\epsfig{figure=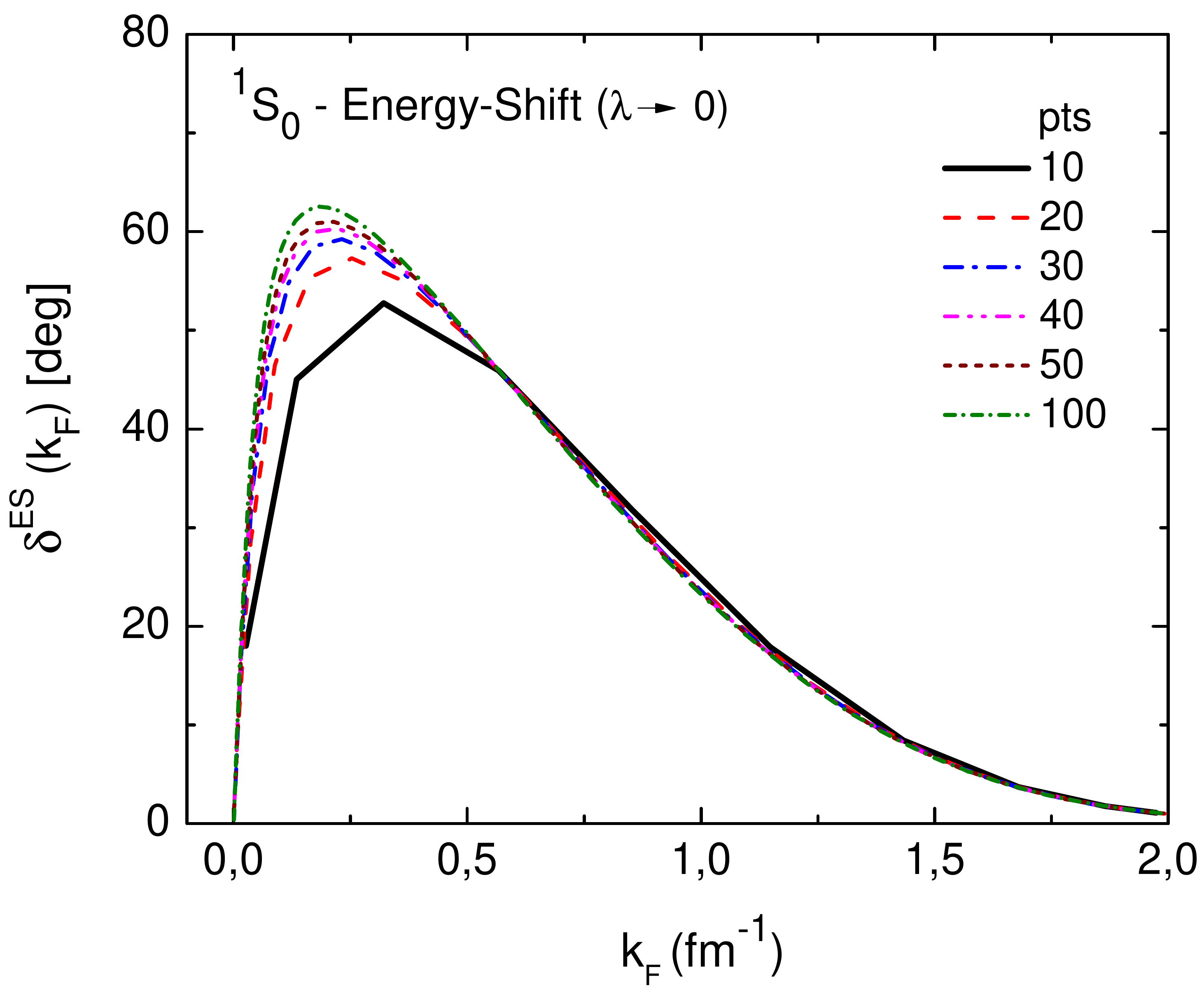,width=0.7\linewidth}
\end{center}
\caption{BCS pairing gap for the $^1S_0$-state in ${\rm MeV}$ (upper panel)
  and the corresponding BCS phase-shift in degrees (lower panel) as a
  function of the Fermi momentum $k_F$ (in ${\rm fm}^{-1}$) in the limit
$\lambda \to 0$. In the upper panel we also compare
  with the SRG invariant phase shift deduced as an energy-shift.
}
\label{fig:pairing-simcut-0}
\end{figure}

In Fig.~\ref{fig:pairing-simcut-0} we show the limiting case $\lambda
\to 0$ where now the pairing gap $\Delta^{\rm ES} (p_n)$ on the grid
is obtained from the SRG-invariant phase-shift computed from the
energy-shift $\delta^{\rm ES} (p_n)$.

The SRG evolution of the BCS pairing gap has been investigated in
Ref.~\cite{Maurizio:2014qsa} for realistic potentials and for the
range of values $\lambda=1-2 ~ {\rm fm}^{-1}$ and for a relatively large
momentum grid. The spread of BCS gap values has been interpreted as a
measure of the uncertainty in the calculation, which we visualize in
our results. However, we see no compelling reason to stop at their
smallest SRG cutoffs besides numerical complications.  As we see,
when we take the limit $\lambda \to 0$, the gap is compatible with
zero in the strict continuum limit.


The on-shell limit has been invoked previously for $A=3,4$ nuclei to
provide an understanding of the Tjon
line~\cite{Arriola:2013gya,Arriola:2016fkr} as well as for unitary
neutron matter~\cite{Arriola:2014tva} to determine the Bertsch
parameter. On a finite grid it is important to check the onset of the
on-shell limit. In many respects this has the properties of a phase
transition where the order parameter is related to the Frobenius
norm (see e.g. \cite{Arriola:2016fkr} for a definition) as we have
suggested recently~\cite{Timoteo:2016vlp}. In other words, for a
finite $\Delta p$ one arrives at an on-shell behavior for $\lambda
\gg \Delta p$. The insensitivity of binding of $A=3,4$ nuclei to the
number of grid points has been explicitly shown in
Refs.~\cite{Arriola:2013gya,Arriola:2016fkr}. In summary, the onset of
the on-shell regime of a discretized momentum system does not require
the strict limit $\lambda \to 0$.

\section{The momentum grid in finite size system}
\label{sec:finit}

Of course, the particular choice of the grid becomes irrelevant for
our considerations, if we take it as a mere auxiliary scheme to solve
the equations numerically. It is interesting, however, to consider grids
corresponding to relevant physical situations, such as a finite size
system as it is the case for finite nuclei. For instance in a
spherical box of radius $R$ the momentum is quantized, $p_n \sim n
\pi/R$, and in this case the spacing in momenta is uniform $\Delta p_n
\sim \pi/R$.

Another way of implementing the finite size of the system is by using
the harmonic oscillator basis with oscillator constant $b=
\sqrt{\hbar/ M \omega}$. Instead of doing this explicitly we will stay
in the momentum basis but tune our discretized momenta to match
harmonic oscillator wave functions.  The equivalence between a
momentum grid and a harmonic oscillator can be obtained from the
differential equation at short distances since
\begin{eqnarray}
-u_{nl}'' + \left[\frac{l(l+1)}{r^2}+ \frac{r^2}{b^4} \right]u_{nl} = p_{nl}^2 u_{nl} \; ,
\end{eqnarray}
so that for $r \ll b$ we have the free wave equation suggesting
\begin{eqnarray}
u_{nl}^{\rm HO}(r) \sim  j_l(p_{nl} r) \; .
\end{eqnarray}
The proportionality constant basis was fixed
in~\cite{kallio1965relation} where
\begin{eqnarray}
p_{nl}^2/2 = (2n+l+3/2)/b^2
\end{eqnarray}
and thus $\Delta p_F \sim 1/b$. For a large volume fermionic system
such as neutron or nuclear matter we can estimate, as usual, the value
of $\hbar \omega$ by matching the total particle number, energy and
radius of closed harmonic oscillator shells to the corresponding
homogeneous Fermi gas in a finite volume. This provides a way of
estimating the weight $w_F$ which corresponds to the momentum spacing
at the Fermi momentum.

For instance, for neutron number $N$, the energy $E$ and the
m.s.r. $\langle r^2 \rangle $ of radius $R$, one gets
\begin{eqnarray}
N &=& 2 \sum_{nl} g_{nl} \to 2 \int \frac{d^3 k d^3 x}{(2\pi)^3}
\theta(k_F-k) \theta(R-r) \; , \nonumber \\ E &=& 2 \sum_{nl} g_{nl} e_{nl}
\to 2 \int \frac{d^3 k d^3 x}{(2\pi)^3} \epsilon_k \theta(k_F-k)
\theta(R-r) \nonumber \\ N \langle r^2 \rangle &=& 2 \sum_{nl} g_{nl}
\langle r^2 \rangle_{nl} \to 2 \int \frac{d^3 k d^3 x}{(2\pi)^3} r^2
\theta(k_F-k) \theta(R-r) \; . \nonumber \\
\end{eqnarray}
Using the accidental degeneracy of the harmonic oscillator $2n+l=k$,
so that $\sum_{nl} g_{nl} = \sum_{k=0}^K (k+1)(k+2)/2 $, we have
\begin{equation}
\epsilon_{nl} = \hbar \omega (2 n + l + 3/2) \equiv \epsilon_k = \hbar
\omega (k+3/2) \; 
\end{equation}
and 
\begin{equation}
\langle r^2 \rangle_{nl} = (b^2 /\hbar \omega) \epsilon_{nl} \; .
\end{equation}
\noindent
In the limit of large $N$ we get $\hbar \omega \sim
1.78~ N^{\frac13}/ (M R^2)$ and $p_F \sim 12.05~ N^{\frac13}/R$ so that
$\hbar \omega \sim 0.147~ k_F /(MR) $ and hence the estimate of the
pairing gap in a large but finite system is
\begin{eqnarray}
\Delta (N,R) \sim \frac{1.78~ N^{\frac13}}{\pi M R^2} \delta \left( \frac{12.05~ N^{\frac13}}{R} \right) \; ,
\end{eqnarray}
or equivalently
\begin{eqnarray}
\Delta_{nn} (k_F,R) \sim \frac{0.147~ k_F}{\pi M R} \delta^{^1S_0}_{nn} \left( k_F \right) \; ,
\end{eqnarray}
\noindent
where $w_F = 0.147/R$. Thus the pairing gap decreases with the size of
the system. Repeating the same exercise for nuclear
matter~\cite{ring2004nuclear} one obtains $\hbar \omega \sim
41~A^{-1/3}~{\rm MeV}$, $b= 1.0 ~ A^\frac16 {\rm fm}$ and $n_{\rm
  max}= 1.58~ A^\frac13$, so that $\Delta p_{n_{\rm max}}^2 = 1/b^2$
and hence since $p_{n_{\rm max}}^2 = 2 M \hbar \omega (n_{\rm
  max}+3/2)$ we have $\Delta p_{n_{\rm max}}= 91~A^{-1/3}~{\rm
  MeV}$. So we have
\begin{eqnarray}
\Delta_{n_{\rm max}} &\sim& \frac{\Delta p_{n_{\rm max}}^2}{2M} \frac{\delta (p_{n_{\rm nmax})}}{\pi} \nonumber\\
&=&\frac{\hbar \omega}{\pi} \delta (p_ {n_{\rm nmax}}) = 13~A^{-1/3} \delta (p_{n_{\rm nmax}}) \; .
\end{eqnarray}

Actually, the corresponding semi-empirical mass formula pairing term
yields a vanishingly small contribution for large $A$ while the size
scales as $R= 1.2~ A^\frac13~{\rm fm}$. Reinstating continuum notation
the final formula reads
\begin{eqnarray}
\Delta(p_F) = \frac{\hbar \omega}{\pi} \delta (p_F) \quad ,  \qquad \hbar \omega
\sim 41 \, A^{-1/3}~{\rm MeV} \, ,
\label{eq:hw-ps}
\end{eqnarray}
where the desired $A^{-1/3}$ can be read off. This is the main consequence of our main result.

For the typical values of the Fermi momentum around $k_F \sim 1~ {\rm
  fm}^{-1}$ the phase-shift presents so linear dependent around
$\delta(p) \sim 30^0$, so that Eq.~(\ref{eq:hw-ps}) can numerically be
approximated by $\Delta \sim 4\, A^{-1/3}~{\rm MeV}$.  This result is
compared to double-differences obtained from stable-nuclei data in
Fig.~\ref{fig:pairing-nuclei}. As we see, and given the fluctuations
in the data, the performance looks of comparable quality as the
textbook pairing term $11 A^{-1/2}~{\rm MeV}$. Note, however, that
after many other effects are explicitly taken into account this liquid
drop model smooth behavior is not confirmed, as discussed in
Section~\ref{sec:mass}. Ours is instead based directly on the
available on-shell $NN$ scattering information relevant for the BCS
pairing gap which not only reproduces the mass number behavior of the
estimated pairing gap from large scale analysis $ \Delta_{\rm exp}
\sim 6(1) \, A^{-1/3}~ {\rm MeV} $ but also accounts by $75 \%$ of the
coefficient. 

These results are rather promising as they use the concept of on-shell
interactions for a finite size system which we regard as the key
aspect of the derivation. However, if we take the discrepancy
seriously we see that there is still room for other effects. In what
follows we consider the possibility that this might be due to
three-body effects with the proviso that the very definition of the
three-body force depends on the definition of the two body force and
more specifically on its off-shell behaviour.  For definiteness we
will analyze this question within the SRG framework.

Of course, while evolving along the SRG trajectory we are on the one
hand keeping the physical phase-shift constant but on the other hand
we are also changing the off-shellness of the interaction striving to
its diagonal on-shell form. The pairing gap is a physical quantity and
we should not expect to depend on unphysical features. From that point
of view, the change in the BCS pairing gap might be viewed as a
genuine uncertainty limiting the predictive power. However, one may
argue that there are three- and many-body forces guaranteeing the
independence of the pairing gap on the calculational
scheme. Unfortunately, within an SRG framework it is unclear at what
SRG scale should the three-body forces be switched on. The prototype
instances for the need of three- and four-body forces are the triton
and helium binding energies. However, it is well known that assuming
some high quality NN potentials as initial condition for the SRG
evolution the need of NNN forces can be minimized by choosing
appropriate
$\lambda$-scales~\cite{Deltuva:2008mv,Jurgenson:2009qs,Jurgenson:2010wy}.
Actually, in Ref.~\cite{Deltuva:2008mv} it is found that using the
CD-Bonn and AV18 (but not N3LO) potentials that the triton binding
energy can be obtained by taking $\lambda \sim 1.5 {\rm fm}^{-1}$.  In
Refs.~\cite{Jurgenson:2009qs,Jurgenson:2010wy} this is confirmed and
in addition the helium binding is obtained also for about the same
$\lambda \sim 1.5 {\rm fm}^{-1}$. Actually, the universal linear
correlation between the binding energies of helium and triton, known
as Tjon line (see e.g. Ref.~\cite{Hammer:2012id} for a review and
references therein), does not depend explicitly on a particular choice
of three-body force; they only relate the equivalent trade-off of the
different off-shellness~\cite{Arriola:2013gya}. This shows that for
some moderate $\lambda$-scales three-body forces are not dominant.

As we have already mentioned the onset of the on-shell NN dynamics
enjoys the features of a phase transition and for a given system the
transition point depends on the momentum resolution and hence on the
number of grid points~\cite{Timoteo:2016vlp}. In that work it was
shown that using the N3LO potential, where $p_{\rm max}= 4 \, {\rm
  fm}^{-1}$ proves sufficient for convergence, the off-shell to on-shell transition
takes place at $\lambda=0.9, 1.1,1.5 \, {\rm fm}^{-1}$ for $N=30,20,10$
respectively. This corresponds to $\Delta p \sim p_{\rm max}/N = 0.13,0.2,0.4 \, {\rm
  fm}^{-1}$ and taking $\Delta p = \pi/R$ to $R \sim 24,16,8 \, {\rm fm}$
respectively. Therefore, for a system of size $R \sim 8 \, {\rm fm}$ the
N3LO becomes on-shell at about $\lambda \sim 1.5 \, {\rm fm}^{-1}$. While
these are rough estimates they show that the scenario in a finite
system where NN forces are close to be on-shell takes place at about
the same scales where three-body forces are not dominating.

Therefore we find that the pairing gap is largely shape independent
and only the strength decreases with the system size. Moreover, we
also see that this feature holds already for moderate SRG cutoffs,
determined by the infrared momentum scale fixed by the finite system
size, $\lambda \sim \Delta p \pi/R$, below which the on-shell regime
sets in. From that point of view we expect finite but not large
corrections to our results stemming from three-body forces.

In the BCS approximation it has been found that the effect of 3N body
forces can effectively be included in the gap equation as a
replacement of the NN interaction in the $^1S_0$
channel as~\cite{Hebeler:2009iv}  
\begin{eqnarray}
V_{NN}(p,k) \to V_{NN} (p,k) + \frac12 \bar V_{3N} (p,k) 
\end{eqnarray}
where the average is understood in the 2-1 relative coordinate below
the Fermi sea. A detailed analysis of this equation along the present 
lines is beyond the scope of our work, and it remains a challenge to 
obtain that, quite generally, for a large $A$ nucleus we may expect 
\begin{eqnarray}
\Delta = ( a_{2N} +a_{3N}+ \dots ) A^{-1/3} + {\cal O} ( A^{-2/3} )
\end{eqnarray}

\section{Conclusions}
\label{sec:conl}

In the present work we have explored the freedom on reducing the
off-shellness of the $NN$ interaction as a way to analyze the BCS
pairing gap. Quite remarkably, we find that there is an on-shell
regime which depends on the size of the nucleus, where the BCS pairing
gap corresponds to the energy shift at the Fermi surface due to the NN
interaction, and can directly determined by the $NN$ phase-shifts in
the $^1S_0$ channel. Our formula provides a satisfactory mass number
dependence $\Delta = 4 ~A^{-1/3}~{\rm MeV}$ which accounts for the
bulk of the $\Delta = 6(1) A^{-1/3}$ found by many nuclear mass
analyses comprising over $2000$ masses and with a mean standard
deviation of $0.2-0.5~{\rm MeV}$. The differences might most likely be
attributed to three-body forces.  This on-shell simplification on the
finite size system neglects $3N$-forces and it remains a challenge to
verify this mass number dependence when they are included in the
calculation. The phenomenological success suggests analyzing more
complicated situations and a thorough analysis of different channels
in higher partial waves. Work along these lines is in progress.

\section*{Acknowledgements}

We thank Artur Polls and Osvaldo Civitarese for informative
discussions.  E.R.A. was supported by Spanish Mineco (grant
FIS2014-59386-P) and Junta de Andaluc\1a (grant FQM225). S.S.  was
supported by FAPESP and V.S.T.  by FAEPEX, FAPESP and
CNPq. Computational power provided by FAPESP grant 2016/07061-3.

\appendix
\section{Toy model}
\label{sec:toy}

In the present study of the $^1S_0$ neutron-neutron channel we use the
toy model gaussian separable potential of the form
\begin{eqnarray}
V(k',k)= C g(k') g(k)
\end{eqnarray}
with $g(k) = e^{-k^2/L^2}$ discussed in our previous
works~\cite{Arriola:2013era,Arriola:2014aia,Arriola:2014fqa,Arriola:2016fkr}
because it allows a handy SRG treatment in the infrared (see
below). For this potential the LS equation, Eq.~(\ref{eq:LS}) can be
solved by assuming $T(k',k) = t g(k') g(k)$ and leading to an explicit
expression for the phase-shift using Eq.~(\ref{eq:phase}), namely
\begin{eqnarray}
 p \cot \delta_0 (p) &=& - \frac{1}{V_0 (p,p)} \left[1- \frac{2}{\pi}
   \dashint_0^\infty dq \frac{q^2}{p^2-q^2} V_0(q,q) \right] \nonumber
 \\ &=& -\frac1{\alpha_0} + \frac12 r_0 p^2 + \dots
\label{eq:ERE}
\end{eqnarray}
where in the last line a low-momentum Effective Range Expansion (ERE)
has been carried out. Parameters $C$ and $L$ have been determined
from the corresponding scattering length $\alpha_0$ and effective range $r_0$ in the
$^1S_0$ channel and the resulting phase-shift is rather reasonable in
the region of CM momentum below $p \sim 1~{\rm
  fm}^{-1}$~\cite{Arriola:2013era,Arriola:2014aia,Arriola:2014fqa,Arriola:2016fkr}.
For this potential the BCS equation is readily solved by taking the
{\it ansatz}
\begin{eqnarray}
\Delta(k) = \Delta_0 g(k)
\end{eqnarray}
where $\Delta_0$ satisfies the implicit equation
\begin{eqnarray}
1 = -\frac1{\pi} \int_0^\infty p^2 dp
\frac{C [g(p)]^2}{M E(p)}
\end{eqnarray}
and $4 M^2 E(p)^2= (p^2-p_F^2)^2 + 4 M^2 \Delta_0^2 g(p)^2$. The
resulting pairing gap $\Delta_F \equiv \Delta_0 g(p_F)$ is depicted in
Fig.~\ref{fig:BCS-toy-vs-pots} and compared to some realistic
potential calculations and, as we see, yields a reasonable
description.



\end{document}